\documentclass[twocolumn,prl,aps]{revtex4}
\usepackage[dvips]{graphicx}

\begin{document}

\draft

\title{Phonon anomalies due to
stripe collective modes in high $T_c$ cuprates}

\author{Eiji Kaneshita, Masanori Ichioka and Kazushige Machida}
\address{Department of Physics, Okayama University,
         Okayama 700-8530, Japan}
\date{\today}

\begin{abstract}
Phonon anomalies observed in various high $T_c$ cuprates by neutron experiments are analyzed theoretically in terms of the stripe concept. The phonon self-energy correction is evaluated by taking into account the  charge  collective modes of stripes, giving rise to dispersion gap, or kink and shadow phonon modes at twice the wave number of spin stripe. These features coincide precisely with observations. The gapped branches of the phonon are found to be in-phase and out-of-phase oscillations relative to the charge collective mode.
\end{abstract}

\maketitle

\narrowtext

%%%%%%%%%%%%%%%%%%%%%%%%%%%%%%%%%%%%%%%%%%%%%%%%%%%%%%%

The pairing mechanism of high $T_c$ superconductors has been still highly debated
since its discovery over a decade ago. The debates are mainly centered on
how to describe the normal state above its superconducting
transition $T_c$. Recently, a remarkable series of neutron experiments on
various cuprates has been done to reveal unequivocally the phonon dispersion anomalies at the particular
reciprocal point.
The phonon anomalies differ slightly in their appearance, ranging from a sharp dispersion jump to
the dispersion kink. This depends on the materials; La$_{2-x}$Sr$_{x}$CuO$_{4}$
(LSCO)\cite{mc,mc1,pinscho}
or YBa$_{2}$Cu$_{3}$O$_{6+x}$(YBCO)\cite{mook,petrov,dogan,mc2} or the focused phonon modes
such as breathing mode, or bond-stretching mode, etc.
This ubiquitous phenomenon for which proper explanation
has not been given so far is interesting because not only it gives a clue to pairing mechanism in high $T_c$ cuprates, but
more importantly it gives a key to better understanding the normal state properties
with full of mysteries.

It is known now that in underdoping La$_{2-x}$Sr$_{x}$CuO$_{4}$ the incommensurate
static magnetic peaks at $\bf Q$
are observed around $(\pi,\pi)$ point in two dimensional CuO$_2$ plane (here the
lattice constant $a$ is unity) by neutron experiments\cite{LSCO}.
As doping proceeds with $x$, the ordering vector $\bf Q$ rotates precisely
at the insulator-metal transition
($x=0.05$) by 45$^\circ$ around $(\pi,\pi)$ point, corresponding to spin modulation
change from the diagonal stripe ${\bf Q}=(2\pi({1\over2}\pm\delta),2\pi({1\over2}\pm\delta))$
to the vertical stripe ${\bf Q}=(2\pi({1\over2}\pm\delta),0)$ or $(0,2\pi({1\over2}\pm\delta))$
where $\delta\sim x$.
In higher doping region the incommesurability $\delta$ almost saturates at a
particular value $\delta\sim {1\over 8}$. The same 45$^\circ$ rotation of the magnetic
superspots is also observed\cite{wakimoto} in (La, Nd)$_{2-x}$Sr$_{x}$CuO$_{4}$, corresponding
to the metal-insulator transition, further lending this superspot rotation
to the universal mechanism
to drive a metal-insulator transition upon doping.
In this material\cite{Nd} and La$_{2-x}$(Sr,Ba)$_{x}$CuO$_{4}$\cite{fujita} the concomitant
static charge superspots at $2\bf Q$ are found by neutron scatterings.

These characteristics coincide nicely with the incommensurate
spin density wave or stripe picture\cite{machida,others} where a solitonic
midgap state accommodates excess holes completely (partially) in an
insulating (metallic) state, giving rise to a microscopic charge and
spin segregation characteristic to topological doping.
The partially occupied midgap state is observed directly
by recent angle-resolved photoemission
 experiment\cite{shen} near $(\pi,0)$ point as a parallel Fermi
surface sheet which is responsible to metallicity of doped CuO$_2$ plane\cite{machida2}.

As for YBCO, although experimental verification of true static magnetic order
comes only from $\mu$SR\cite{musr} which reports a similar phase diagram
to LSCO
in doping vs. static magnetic ordering, there are several common characteristics
mentioned above\cite{note}, including neutron scatterings\cite{mook2}.
Thus it is expected that the stripe picture may be applicable to YBCO\cite{note2}.
Numerous other experiments  such as wipe-out effect for charge order by
NMR\cite{imai}, detailed transport measurements\cite{ando}, etc. and theories\cite{sorella},
collectively point to support the idea that the stripe concept may be applicable to
doped CuO$_2$ plane in general.

Here we are going to
demonstrate that the phonon anomalies observed in various cuprates
are  a direct manifestation of existence of the low-lying collective charge excitations
associated with stripes. Physically, the $\pi$-phase shifting
domain walls accommodate excess holes and
the remaining antiferromagnetic regions are intact. The domain wall motion, or
the charge collective oscillation,
is strongly coupled with the particular phonon mode, such as breathing or bond-stretching modes. Both electron and phonon systems are mutually influenced by the others to lower the total energy.
The spin and charge inhomogeneity is a universal phenomenon in
doped CuO$_2$ plane, out of which high $T_c$ superconductivity is created. This process is still beyond our understanding at the moment, but this identification, we believe, is important towards this step.
This is particular true for YBCO where there is no defining experiment to uncover the charge collective modes which has been elusive so far.
Since the stripe concept, although originally derived from mean field approximation\cite{machida,others}, may be valid for describing a doped Mott insulator, or at least, the stripe is a convenient ``vehicle'' to accommodate excess holes, it is worthwhile to pursue this identification.

We start with the Hubbard model $H=\Sigma_{i,j,\sigma}t_{i,j}c^{\dagger}_{i,\sigma}
c_{j,\sigma}$ $+U\Sigma_{i}n_{i\uparrow}n_{i\downarrow}$ to describe a stable stripe
and its excitations. The mean-field ground state\cite{machida,machida2}
and random phase approximation
(RPA)\cite{kaneshita} yield a rich structure for individual and collective excitation spectra
of spin and charge channels; namely the dynamical transverse (longitudinal) susceptibility
$\chi_{xx(zz)}({\bf q},\omega)
=\langle\langle S_{x(z)};S_{x(z)}\rangle\rangle_{{\bf q},\omega}$,
and the dynamical charge susceptibility
$\chi_{nn}({\bf q},\omega)=\langle\langle n;n\rangle\rangle_{{\bf q},\omega}$.
The RPA equation for $\chi_{S_\uparrow S_\downarrow}
= \langle\langle S_\uparrow; S_\downarrow \rangle\rangle$ is written as
$\chi_{S_\uparrow S_\downarrow}({\bf r}_1,{\bf r}_3, \omega)$
$=\chi_0^{\uparrow \downarrow} ({\bf r}_1,{\bf r}_3, \omega)$
$+ U \sum_{{\bf r}_2} \chi_0^{\uparrow \downarrow}
({\bf r}_1,{\bf r}_2, \omega)$
$\chi_{S_\uparrow S_\downarrow}({\bf r}_2 ,{\bf r}_3 , \omega)$
with non-interacting susceptibility $\chi_0^{\uparrow \downarrow}$.
After Fourier transformation to ${\bf k}$-space,
this is reduced to a matrix equation of
$\chi_0^{\uparrow \downarrow}$ and $\chi_{S_\uparrow S_\downarrow}$.
By solving it, we obtain
$\chi_{xx}({\bf q}+l_1 {\bf Q},{\bf q}+l_2 {\bf Q}, \omega)$.
There, ${\bf q}+l_1 {\bf Q}$ and ${\bf q}+l_2 {\bf Q}$ are, respectively,
outgoing and incoming wave vectors.
They can change by $(l_1-l_2) {\bf Q}$ through the Umklapp process,
since the underlying electronic state is spatially modulated with
the periodicity ${\bf Q}$.
In a similar manner, we calculate
$\langle\langle n_\uparrow;n_\uparrow \rangle\rangle$ and
$\langle\langle n_\uparrow;n_\downarrow \rangle\rangle$, and obtain
$\chi_{zz}({\bf q}+l_1 {\bf Q},{\bf q}+l_2 {\bf Q}, \omega)$
and $\chi_{nn}({\bf q}+l_1 {\bf Q},{\bf q}+l_2 {\bf Q}, \omega)$.

\begin{figure}
\includegraphics{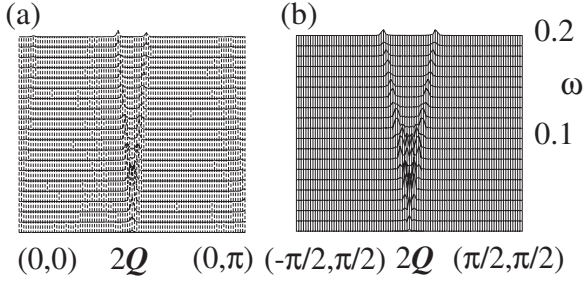}
\caption{
Dispersion curves of the charge collective mode around $2{\bf Q}$.
We plot  Im$\chi_{nn}({\bf q}, \omega)$  for two paths along $2{\bf Q}$ (a) and perpendicular to $2{\bf Q}$ (b).
}
\label{fig:charge}
\end{figure}

The spin modes $\chi_{xx(zz)}({\bf q},\omega)\equiv\chi_{xx(zz)}({\bf q},{\bf q},\omega)$
give rise to collective excitation at the ordering vector $\bf Q$ and its odd-harmonic points
as a Goldstone mode.
The charge mode $\chi_{nn}({\bf q},\omega)\equiv\chi_{nn}({\bf q},{\bf q},\omega)$ exhibits
the collective excitations at ${\bf q}=2{\bf Q},4{\bf Q}\cdots$.
There, $\chi_{xx}$ gives spin wave,
when we assume that the magnetization points to the $z$ axis.
The translational phason modes appear in $\chi_{nn}$ ($\chi_{zz}$),
which correspond to the motion of the charge (spin) stripe\cite{kaneshita}.
Among various individual and collective motions associated with
stripes, the phason mode in the charge oscillation is directly
coupled to the phonons and most relevant to the phonon
renormalization.
In Fig.1, ${\rm Im}\chi_{nn}({\bf q},\omega)$ is shown
around ${\bf q}=2{\bf Q}=(0, \frac{\pi}{2})$ along the $q_y$- and the
$q_x$-directions.
This particular charge
mode with the anisotropic velocities is situated at $2{\bf Q}$.
The motion parallel (perpendicular) to $2{\bf Q}$ vector corresponds to
the compression (meandering) motion of stripes with high (low) velocity.
In the following we illustrate typical results for the metallic vertical
stripe \cite{machida2} which is stable with ${\bf Q}=(\pi,\frac{3\pi}{4})$ for
the electron filling per site $n=0.82$
and the parameters are fixed as ${U\over t}=4.0$ and ${t'\over t}=-0.2$
($t (t')$ is the (next) nearest neighbor hopping integral).
The main features of the results do not much depend on the choice
of these values. We note that $t$ is an order of 0.5eV and the phonon energy
$\omega\sim 80$meV for the breathing mode or bond-stretching mode which
is coupled strongly to the electronic state. We focus on an energy range of typically ${\omega\over t}\sim0.1$
in the following. All the energy is scaled by $t$.

\begin{figure}
\includegraphics{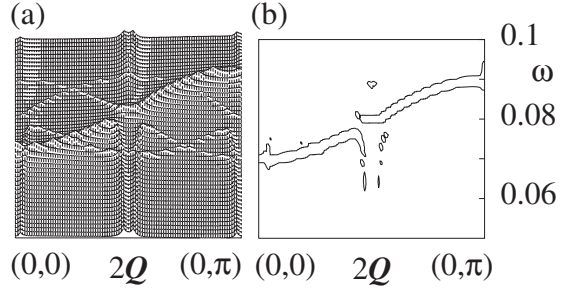}
\caption{
Renormalized phonon spectral functions for  $|g|^2=0.001$.
(a) is logarithmic plot to emphasize the small intensity of the shadow bands.
(b) is contour plot shown for high intensity.
The simple sinusoidal form is assumed for
the unperturbed phonon ($A=-0.01,B=0.08,k=0$)\protect\cite{dis}.
}
\label{fig:sine}
\end{figure}

In order to study the effects of the stripes on the phonon Green's function
$D({\bf q}+l_1{\bf Q},{\bf q}+l_2{\bf Q},i\omega_n)$
within RPA, we consider the following standard electron-phonon Hamiltonian:
$H'=\Sigma_{{\bf k},{\bf q},\sigma}g
c^{\dagger}_{{\bf k}+{\bf q},\sigma}c_{{\bf k},\sigma}(b_{\bf q}+b^{\dagger}_{-{\bf q}})$
with $g$ being the coupling constant. Since the electron bubble in the phonon
self-energy correction is nothing but the dynamical charge susceptibility mentioned,
we immediately obtain an equation in a matrix form for the thermal Green's functions
$\Sigma_{m'}\{\delta_{lm'}-|g|^2 D_0({\bf q}+l{\bf Q},i\omega_n)
\chi_{nn}({\bf q}+l{\bf Q},{\bf q}+m'{\bf Q},i\omega_n) \}
D({\bf q}+m'{\bf Q},{\bf q}+m{\bf Q},i\omega_n)=D_0({\bf q}+l{\bf Q},i\omega_n) \delta_{lm}$,
where the unperturbed Green's function
$D_0({\bf q},i\omega_n)=-2\omega_{\bf q} /(\omega^2_n+\omega^2_{\bf q})$,
and $\omega_{\bf q}$ is the unperturbed phonon dispersion.
Diagonalization of this matrix equation yields the renormalized
phonon Green's function.

\begin{figure}
\includegraphics{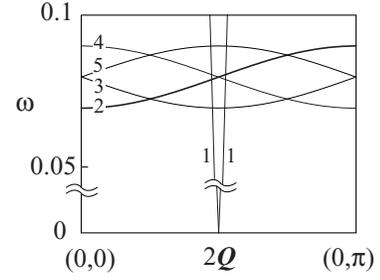}
\caption{
Schematic figure for relevant modes. 1: Charge collective mode. 2: Unperturbed phonon. 3,4, and 5: Phonons shifted by 2${\bf Q}$, 4${\bf Q}$ and 6${\bf Q}$ via Umklapp processes.
}
\label{fig:schema}
\end{figure}

We show the renormalized phonon spectral function
$-{1\over \pi}{\rm Im}D({\bf q},{\bf q},i\omega_n\rightarrow\omega+i\eta)$ in Fig.2,
where we assume a simple sinusoidal form\cite{dis} for $\omega_{\bf q}$ as a typical phonon.
It is seen from Fig.2 that the dispersion exhibits several characteristic modifications:
(A)  The discontinuities of the dispersion occur at various points, notably near $2{\bf Q}$
where the dispersion cone of the collective charge mode shown in Fig.1 and
the phonon mode intersect.
(B) The appearance of the downward hybridized
new phonon dispersions.
(C) The flat dispersion appears above the gap at $2{\bf Q}$.
(D) The shadow band which is a replica of the original mode is seen.

These features come from the two physical reasons schematically illustrated. in Fig.3:
The hybridization of the charge collective mode centered at $2{\bf Q}$ denoted as
1 with the original phonon 2 in Fig.3 and the band-folding due to
the periodic structure with the charge ordering vector $2{\bf Q}$, which amounts to
shifting the original dispersion 2, resulting in phonon dispersions; curve 3 $\sim$5
in Fig.3. The hybridization of these dispersions gives rise to the reconnected phonon modes in Fig.2.

These main four features (A-D) of the dispersion modification become more
evident when the electron-phonon coupling $|g|^2$ increases,
or the unperturbed dispersion is situated at lower energy side.
The gap size at the discontinuity increases as the original phonon band is widen.
It is noteworthy that the discontinuity of the dispersion, or a kink feature and associated new lower mode
branching-out from the disconnected lower phonon  are reminiscent of the observation
on  O$_{6.35}$ and O$_{6.60}$ in YBCO by Mook and Do\u{g}an\cite{mook} (see
their Figs.1(d) and (e)).
Note also that since the discontinuity point occurs at
the intersection of the charge collective mode at $2{\bf Q}$ and phonons,
the observed dispersion anomaly should change with doping;
As doping increases the $2{\bf Q}$ vector increases.
This is indeed observed by neutron experiment\cite{dogan} and also consistent with the doping dependence of the fundamental vector ${\bf Q}$ of the magnetic peaks\cite{mook2}. These constitute an internal consistent picture for the stripe origin.

\begin{figure}
\includegraphics{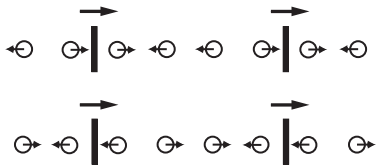}
\caption{
Schematic figure for two in-phase (lower energy) and out of phase
(higher energy) oscillation patterns corresponding to the two branches above and below the gap at $2{\bf Q}$.
The vertical bars show the charge domain walls of the stripes.
}
\label{fig:osci}
\end{figure}

The physical implication of the phonon gap formation is understood by analyzing the phonon oscillation  patterns of the upper and lower phonon branches.
According to linear response theory, the oscillation patterns  are given by
$\phi({\bf r},t)=\Sigma_l{1\over\hbar}D({\bf q}+l{\bf Q},{\bf q},\omega)e^{il{\bf Q}\cdot {\bf r}}h_{{\bf q}}e^{i{\bf q}\cdot{\bf r}-i\omega t}$
%$$
%\phi({\bf r},t)=\Sigma_l{1\over\hbar}D({\bf q}+l{\bf Q},{\bf %q},\omega)e^{il{\bf Q}\cdot {\bf r}}h_{{\bf q}}e^{i{\bf q}\cdot{\bf %r}-i\omega t}
%$$
%\noindent
where $\phi({\bf r},t)$ is the lattice displacement and $h_{{\bf q}}$ is the
external field with the wave number ${\bf q}$ and the frequency $\omega$.
At the wave number $2{\bf Q}=(0, \frac{\pi}{2})$,
the oscillations corresponding to the upper ($\omega=0.08$) and
lower ($\omega=0.075$) branches shown in Fig.2 are schematically displayed in Fig.4. The four site period oscillations are either in phase or out of phase
with the charge oscillation in the electron system which is the phason motion of the stripes.

\begin{figure}
\includegraphics[bb=174 425 384 629]{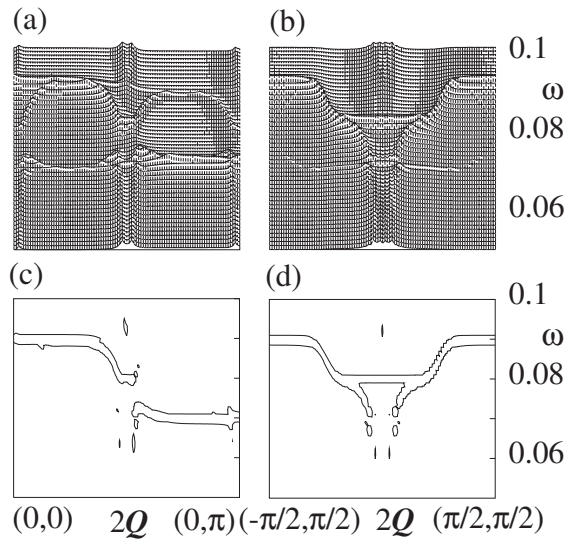}
\caption{
Renormalized phonon spectral functions for two perpendicular paths. $|g|^2=0.001$.
(a) and (c) $(0,q_y)$ for $0\leq q_y\leq \pi$. (b) and (d) $(q_x,{\pi\over2})$ for $|q_x|\leq {\pi\over2}$.
The unperturbed phonon with a steep slope ($A=0.01,B=0.08,k=0.9999$)\protect\cite{dis} is assumed. (a) and (b) are logarithmic plots. (c) and (d) are contour plots.
}
\label{fig:sn}
\end{figure}

We show the renormalized dispersions parallel to $2{\bf Q}$ in Figs.5(a) and (c) and perpendicular
to $2{\bf Q}$ centered at $2{\bf Q}$ in Figs.5(b) and (d). Here we assume
the unperturbed phonon with a steep slope at $(0,{\pi\over 2})$, simulating  the breathing mode in LSCO with the energy $\sim$80meV.
Noticeable features of the modification are:
(A) Rather large dispersion discontinuity observed at $2{\bf Q}$ is seen in Figs.5(a) and (c).
(B) It is noted that the shadow mode which is a replica of the
original band can be seen in Figs.5(a) and (b).
(C) The anomaly continues to extend to the perpendicular direction seen
from Figs.5(b) and (d).
These features (A-C) coincide with the neutron experiment\cite{mc} on LSCO with
$x=0.15$ where at ($ \frac{\pi}{2},0)$ in the $(q_x,0)$ scan they observe a sharp
drop of the bond-stretching mode (see their Fig.2) and this anomalous
behavior continues to the perpendicular direction (see their Fig.4).
As they emphasize, the phonon anomaly occurs not
because of the simple band-folding effect due
to new charge periodicity with 4 lattice period which
would result in a gap at the new Brillouin zone boundary at $({\pi\over4},0)$,
but the observed gap is at $({\pi\over2},0)$.

The anomalous reciprocal regions for the two perpendicular directions have different widths as seen from Fig.5 because the underlying
charge cone  is anisotropic as mentioned, yielding
different widths (compare the openings of the cones at the center in Figs.1(a) and (b)
corresponding to the cones in Figs.5(a) and (b)). Thus it is possible to use it as a kind of spectroscopy, namely, to investigate this
anisotropy of the charge collective mode through the analysis
of the phonon anomaly. In fact according to the above experiment\cite{mc} the width of
the anomalous region in the perpendicular direction
is wider than that in the parallel direction. This accords with our result in Fig.5. We can attribute it as arising from the lower velocity of the meandering mode than the compression mode\cite{kaneshita}.

McQueeney, {\it et al.}\cite{mc2} report that there is no phonon anomaly below $x\leq0.05$
in LSCO where the  static diagonal stripe is stabilized.
There are two possibilities: (1) Since the velocity of
the collective charge excitation is rather low for the insulating diagonal stripe\cite{kaneshita}, the phonon mode investigated by them may not be
coupled with this excitation, failing to drive the anomaly. (2) As the second possibility, the collective charge excitation could be gapped\cite{kaneshita}. There is no low-lying excitation coupled to phonons.
In this connection,
we point out that in Cr, where the diagonal stripe in the present terminology
is realized, neutron experiment to see the phonon anomaly at 2${\bf Q}$
corresponding to the so-called strain wave or charge order is desirable.

Considering the electron-phonon coupling, we study the phonon anomalies in
the stripe state. The phonon anomalies observed mainly
by neutron scattering experiments on LSCO and YBCO with various
doping levels are accounted for as arising from two physical effects: Hybridization with the charge collective mode at $2{\bf Q}$ and
band-folding of a phonon band. These give rise to reconnection and gap formation in bands.
The underlying physics behind the phonon anomalies at the particular
reciprocal point, namely  $2{\bf Q}$, is the existence of the collective charge excitations associated with stripes;
the translational phason and meandering gapless modes with different velocities situated at every even-multiples
of the fundamental vector ${\bf Q}$ which varies systematically with hole-doping.
Thus this also leads to systematic changes of the $2{\bf Q}$ vector, a fact that the experiments
have found (see Fig.4 in Ref.\cite{dogan}).
In order to confirm our identification, further neutron experiments are desirable, in
particular to find the yet-uncovered lower modes branching-out downwardly
from the phonon discontinuity
at $2{\bf Q}$. Since in principle the phonon anomalies occur for all phonon modes and
the observability depends on the electron-phonon coupling constant,
lower phonon modes are more favorable for this purpose.

It might be interesting to point out that similar phonon mode anomalies are also
found in (La,Sr)$_2$NiO$_4$\cite{tranquada}
where the diagonal stripes are stabilized because as mentioned
this system is an insulator up to higher dopings, thus 45$^{\circ}$ rotation of the superspots occurs.
In yet another non-cuprate superconductor Ba$_{0.6}$K$_{0.4}$BiO$_{3}$,
which is related to charge density waves instead of spin density wave here
the phonon anomaly of the
bond-stretching mode appears at a certain reciprocal point\cite{braden}.
We believe that these anomalies could be analyzed in the same way as in here,
which belongs to future problem.

We thank J. Tranquada, R. McQueeney H. Mook and N. Wakabayashi for their useful discussions
and  information.

%%%% references %%%%%%%%%%%%%%%%%%%%%%%%%%%%%%%%%%%%%%%

%%%%%%%%%%%%%%

%%%%%%%%%%%%%%%%

\begin{references}
\bibitem{note}
It is noteworthy that by applying stress the intensities of four incommensurate
spots near $(\pi,\pi)$ changes into two pair of spots with unequal intensity,
implying a one-dimensional stripe originally envisaged theoretically\cite{machida,others}.
\bibitem{note2}
Note that static charge order is observed in YBCO$_{6.45}$
which is referred to in Ref.\cite{dogan}.
\bibitem{dis}
We assume the dispersion for the unperturbed phonon:
$\omega_{{\bf q}}=-{A\over2}(1-sn(q_x+K))sn({q_y\over q_x}2K)
+{A\over2}(1+sn(q_x+K))+B$ for $|q_x|\geq|q_y|$
and $q_x\leftrightarrow q_y$ for $|q_x|<|q_y|$
where $sn(q)$ is the elliptic function and $K$ complete elliptic integral
with modulus $k$.
\end{references}
\end{document}